\documentclass{aastex}
\usepackage{emulateapj5}
\usepackage{amsmath,natbib}

\begin{document}

\title{Bardeen-Petterson Effect and Quasi-periodic Oscillations in X-Ray Binaries}

\shorttitle{Bardeen-Petterson Effect in X-Ray Binaries}

\author{P. Chris Fragile, Grant J. Mathews, and James R.
Wilson\altaffilmark{1}}

\affil{Center for Astrophysics, University of Notre Dame, Notre
Dame, IN 46556}

\email{pfragile@nd.edu}
\email{gmathews@nd.edu}
\email{jimwilson@llnl.gov}

\altaffiltext{1}{Lawrence Livermore National Laboratory, Livermore, CA  94550}

\newfont{\ms}{cmsy10 at 12pt}


\begin{abstract}
The Bardeen-Petterson effect around a rapidly-rotating compact
object causes a tilted accretion disk to warp into the equatorial
plane of the rotating body.  Viscous forces cause the accretion
flow to divide into two distinct regions - an inner aligned
accretion disk and an outer tilted accretion disk.  The transition
between these two occurs at a characteristic radius that depends
on the mass and angular momentum of the central object and
possibly on the accretion rate through the disk.  We propose that
accreting material passing through the transition region may
generate quasi-periodic brightness oscillations (QPOs) such as have
been observed in a number of X-ray binaries.  We show that this
effect may be present in the
black-hole X-ray binary GRO J1655-40.  We also argue that the QPO frequency range predicted by this model is consistent with observed QPO frequencies in both black-hole and neutron-star low-mass X-ray binaries.
\end{abstract}


\section{Introduction}
Since their discovery in the mid-1980s \citep{van85},
quasi-periodic oscillations (QPOs) have been observed in the X-ray
brightness of a number of neutron-star (NS) and black-hole (BH)
X-ray binaries.  In so-called $Z$-sources, which are
high-luminosity NS low-mass X-ray binaries (LMXBs), there are
typically four distinct QPO frequencies observed: 1) The $\simeq5-20$
Hz normal/flaring branch oscillation; 2) The $\simeq15-60$ Hz
horizontal-branch oscillation; 3) \& 4) Two ``kilohertz'' QPOs in the
range $\simeq200-1200$ Hz \citep{van96}.  Atoll sources, which
are less luminous NS LMXBs, typically display two kHz QPOs
in the range $\sim500-1330$ Hz as well as low frequency ($\sim20-60$ Hz)
QPOs and broad noise components \citep{str96}.  Not as many detections exist for QPOs in BH
X-ray binaries.
However, several low-frequency ($\sim0.1-50$ Hz) QPOs have been
observed, and higher-frequency QPO components have been identified in the spectra of at least three BH candidates [$\simeq 67$ Hz in GRS
1915+105 \citep*{mor97}; $\simeq 300$ Hz in GRO J1655-40
\citep{rem99b}; and $\sim 160-280$ Hz in XTE J1550-564
\citep{rem99a,hom00}].  (See, e.g. \citealt{van95}, \citealt{van00b}, and \citealt{wij00} for recent summaries.)

These QPOs may provide a useful probe into the inner accretion
flows around neutron stars and black holes.  Consequently, a
number of theoretical models have been proposed to explain the
observations. For example, QPOs have been modeled as general
relativistic Lense-Thirring precession \citep*[e.g.][]{ste98,mor99}, geodesic motion of gas clumps \citep*[e.g.][]{ste99a,ste99b,kar99}, trapped oscillation
modes in the disk \citep*[e.g.][]{mar98,now91,now92},
interruption of the gas flow at the centrifugal barrier \citep*[e.g.][]{tit98}, and various beat-frequency interpretations
(e.g. \citealt*{alp85}; \citealt{lam85}; \citealt*{mil98}).  Some
of these models depend explicitly on the existence of a stellar
surface and/or a large-scale magnetic field and are therefore only valid
for QPOs in NS systems.  Furthermore, none of the models proposed thus far has been able to simultaneously explain all of the QPOs observed in individual systems.

Most recent theoretical attention has focused on the high-frequency kHz QPOs.  In this paper, however, we wish to instead address moderate-frequency ($\sim 1-300$ Hz) QPOs.  Specifically, we propose the Bardeen-Petterson effect \citep*{bar75} as a potential mechanism for
generating such QPOs in X-ray binaries.  This new mechanism derives from the possible
existence of a well-defined transition radius in tilted accretion
disks around rapidly-rotating compact objects.  This transition
radius is a result of the Bardeen-Petterson effect
which will be explained further in Section \ref{BPsect}.  In Section \ref{XBsect} we
consider the likelihood that this effect is present in BH and NS
X-ray binaries.  We also discuss mechanisms by which
accreting material passing through this transition radius may
generate observable quasi-periodic brightness oscillations.  In
Section \ref{GRO J1655-40 sect} we discuss the BH X-ray binary GRO
J1655-40. We find that our estimated QPO frequency range is consistent with wither of two QPOs observed in that system.  In Section \ref{LMXB sect} we consider neutron-star LMXBs and argue that our model is consistent with the $\sim 100$ Hz peaked-noise component seen in some of these systems.

\section{The Bardeen-Petterson Effect} \label{BPsect}
\begin{figure}
\epsscale{0.5}
\plotone{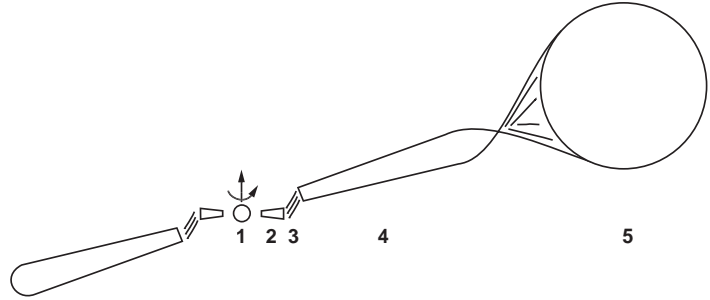}
\caption{Schematic diagram of the Bardeen-Petterson effect showing: (1) the central rotating black hole or neutron star; (2) the inner, aligned accretion disk; (3) the transition region; (4) the outer, tilted accretion disk; and (5) the companion star.\label{BPEffect}}
\end{figure}

The Bardeen-Petterson effect is illustrated schematically in Figure \ref{BPEffect}.  This effect causes a tilted accretion disk around
a rapidly-rotating compact object to warp into the equatorial
plane of the central object \citep*{bar75,kum85,sch96}.  The effect is the
combined result of differential Lense-Thirring precession and
internal viscosity.  The Lense-Thirring precession causes the disk
to ``twist up.''  Damping of the fluid motion by viscosity
limits this twisting.  Close to the accreting black hole or neutron
star, where the Lense-Thirring precession is strongest, viscosity
allows the misaligned angular momentum of the tilted disk to be
transported outward.  This allows the inner region to settle into
the rotation plane of the BH or NS.  Further out, the disk remains
in its original plane because the Lense-Thirring precession rate
drops off rapidly with increasing radial distance [$\sim r^{-3}$ \citep{len18}]. Internal pressure and
viscous stress in the outer regions are thus able to limit its
effects. The end result, as shown in Figure \ref{BPEffect}, is an aligned inner accretion disk (region 2), a
tilted outer accretion disk (region 4), and a transition region (3) between the two.

\subsection{Transition Radius}
The Bardeen-Petterson transition radius is expected to occur
approximately where the rate of twisting up by differential
precession is balanced by the rate at which warps of the disk are
diffused or propagated away by viscosity.  Using thin disk theory,
in which the effects of pressure are neglected and an isotropic
viscosity is assumed, the transition radius was first estimated as
\citep*{bar75,hat81}:
\begin{equation}
R_{\mathrm{BP}} = \left[ \frac{6 a_*}{\alpha \delta^2} \right]^{2/3} R_{\mathrm{GR}}~.
\end{equation}
Here $\alpha$ is the standard viscosity parameter associated with
the radial accretion \citep*{sha73}, $\delta$ is the aspect ratio
of the disk, $a_*$ is the dimensionless specific angular momentum parameter of the BH or NS ($a_*=Jc/GM^2$ where $J$ is the angular momentum), and $R_{\mathrm{GR}}=GM/c^2$, where $M$ is the gravitational mass of the BH or NS.

\citet*{iva97} obtained an alternative expression for a characteristic radius
analogous to the Bardeen-Petterson transition radius by
considering the scale of oscillations in a twisted accretion disk
around a Kerr BH:
\begin{equation}
R_{\mathrm{BP}} = \left[ \frac{384 a_*}{25 \delta^2}\right]^{2/5} R_{\mathrm{GR}}~.
\end{equation}

Most recently, \citet*{nel00} obtained a slightly different estimate by
considering separate horizontal and vertical shear components of
viscosity.  Their expression for the transition radius depends on the relative strength of the
dimensionless vertical shear viscosity parameter, $\alpha_1$:
\begin{equation}
R_{\mathrm{BP}} = 
\begin{cases}
\left[ \frac{24\alpha_1 a_*}{\delta^2}\right]
^{2/3}R_{\mathrm{GR}}~, & \text{if $\alpha_1>\delta$,} \\
\left[ \frac{24 a_*}{\delta}\right]
^{2/3}R_{\mathrm{GR}}~, & \text{if $\alpha_1\lesssim\delta$.}
\end{cases}
\label{NPeq}
\end{equation}

Typical values for the viscosity parameters
$\alpha$ and $\alpha_1$, and the aspect ratio $\delta$, are $0.04\lesssim\alpha\lesssim0.2$, $\alpha_1\sim\alpha$, and
$0.03\lesssim \delta\lesssim0.3$.  These are based on numerical calculations \citep[e.g.][]{nel99,bry99}.  In
hydrodynamic simulations \citet*{nel00} found the transition
radius to lie consistently below the analytic estimates in
equation \ref{NPeq} by a factor of $\sim2-3$.  This correction is included in the following analysis,
although it does not qualitatively change the results.

In the present work we will adopt the following expression for the
Bardeen-Petterson transition radius motivated largely by equation \ref{NPeq} and the numerical results of \citet{nel00}:
\begin{equation}
R_{\mathrm{BP}}=A a_*^{2/3} R_{\mathrm{GR}}~,
\label{RBP}
\end{equation}
where the scaling parameter $A$ lies in the range $10\lesssim A\lesssim300$.  The corresponding Keplerian orbital frequency is therefore:
\begin{equation}
\nu_{Kep,\mathrm{BP}}=\frac{c^3}{2 \pi G A^{3/2} a_* M}~.
\label{fBP}
\end{equation}

\section{Bardeen-Petterson Effect in X-Ray Binaries} \label{XBsect}
As noted above, the Bardeen-Petterson effect requires a tilted
accretion disk around a rapidly-rotating compact object.  In the
context of our present study, we must evaluate the likelihood that
this condition is met in observed X-ray binaries.  To do this,
we must consider the likelihood that tilted-disk
systems are formed and how long they last relative to the overall lifetime of
X-ray binaries.  These factors are not well known, but we can make some estimates.

\subsection{X-Ray Binary Formation}
X-ray binaries can possibly form in one of four ways
\citep*[e.g.][]{can90,ver95}.  The most likely paradigms for Galactic sources are core-collapse of a massive star in a
binary or accretion induced collapse of a white dwarf or neutron star in a binary.  Less likely possibilities (except perhaps in dense globular clusters) are binary capture or binary replacement.  In the first two
mechanisms one member of a preexisting binary must explode as a supernova in order to form the required NS or BH.  During this violent event even a small asymmetry in the explosion will impart substantial momentum to the newly-formed remnant \citep[e.g.][]{dew87,bai89}.  Even if an initially aligned system remains bound, it is unlikely that the two stellar spin axes will remain aligned with the orbital spin axis after the supernova.  Detailed Monte Carlo modeling by \citet{bra95} of this evolutionary path for LMXBs indicates that only $\sim 20$\% of the progenitor systems remain bound after a supernova explosion.  Furthermore, the resultant median of tilt angle for the systems that remain bound is approximately $20^\circ$, with roughly 60\% of them having tilt angles in the range of $5^\circ - 45^\circ$.  Hence, a large fraction ($\sim 1/2$) of such systems should be born as candidates for the Bardeen-Petterson effect.

In the other two mechanisms, which are unlikely except perhaps in globular clusters, a NS or BH must pass close enough to another star or binary system to become gravitationally bound.  Here the collapsed star and its new companion evolved separately until the time of their encounter.  There is therefore no preferred orientation for the angular momenta of the stars relative to the angular-momentum axis of the forming binary.  Hence the majority of such systems would also be Bardeen-Petterson effect candidates.

Thus, it appears likely that many, if not most, X-ray binaries form with misaligned angular momenta (one such observed system is discussed in Section \ref{GRO J1655-40 sect}).

\subsection{Alignment Timescale}
Since the Bardeen-Petterson effect causes the accretion flow to lie in the equatorial plane of the accreting body out to a disk radius of $R_\mathrm{BP}$, the material actually accreted by the stellar remnant will not change the orientation of the remnant's angular momentum.  However, as pointed out by \citet{ree78}, the torque exerted by the BH or NS as it aligns the disk with its own spin axis also has the effect of aligning the BH or NS with the angular momentum of the disk.  A number of authors have addressed this issue \citep*[e.g.][]{ree78,sch96,nat98}, particularly with respect to misaligned accretion disks around black holes.  The alignment angle decreases exponentially with a characteristic timescale, $t_{align}$, given by \citep{sch96}:
\begin{equation}
t_{align}\approx 3 a_* \frac{M}{\dot{M}}\left( \frac{2R_{\mathrm{GR}}}{R_{\mathrm{BP}}}\right)^{1/2}~,
\end{equation}
where $\dot{M}$ is the mass accretion rate through the disk.  For illustration, we consider a 1.4 $M_\odot$ NS and a 7 $M_\odot$ BH, each with
$a_* =0.1$ and $A=100$.  We derive a lower limit for $t_{align}$ by setting $\dot{M}$ at the respective Eddington accretion limits.
These values give minimum alignment timescales of $3 \times
10^6$ yr and $7 \times 10^6$ yr, respectively.  However, a more realistic long-term accretion rate for LMXBs is $\dot{M}=0.01 \dot{M}_\mathrm{Edd}$ \citep{van95}.  This gives alignment timescales comparable to the estimated lifetime of X-ray binaries [$10^8-10^9$ yr \citep{ver95}].  These long timescales for alignment imply that the typical distribution of tilt angles in LMXBs should be close to the distribution estimated by \citet{bra95} for LMXB formation.  Thus, a large fraction ($\sim 1/2$) of LMXBs may be candidates for observation of the Bardeen-Petterson effect.

\subsection{X-Ray Modulation Mechanism} \label{Mechsect}
We now wish to consider possible mechanisms by which the Bardeen-Petterson effect could generate observable QPOs.  We begin with the findings of \citet*{nel00} that for moderate angles ($\theta \gtrsim 30^\circ$) between the disk mid-planes, the inner and outer accretion disks are nearly disconnected.  For larger angles, the separation between the two disks becomes even more pronounced.  In such cases, accretion between the outer and inner disks presumably occurs along a tenuous bridge of material or in the form of gas clumps that periodically break off from the outer disk and collide with the inner disk.  Whether accretion takes place along a bridge or in discrete clumps, there is likely to be significant shock heating at the point where this accreting material impacts the inner disk.  As the shock-heated gas continues to orbit the BH or NS, it may generate periodic brightness oscillations at the orbital frequency for that radius.  We expect this radius to lie just inside the Bardeen-Petterson transition radius, so that the QPO frequency should be very close to the Keplerian orbital frequency given in equation \ref{fBP}.  Using $\theta \gtrsim 30^\circ$ as a guide, we may expect $\sim 30$\% of LMXBs to be susceptible to this mechanism based on the tilt-angle distribution predicted by \citet{bra95}.

\section{GRO J1655-40} \label{GRO J1655-40 sect}
The black-hole binary GRO J1655-40 is a very important target for
studies such as this one since many of the relevant system parameters have been measured or inferred.  There is also evidence that the angular-momentum axis of the black hole in this system is not aligned with the angular momentum of the binary.

\begin{figure}
\epsscale{0.5}
\plotone{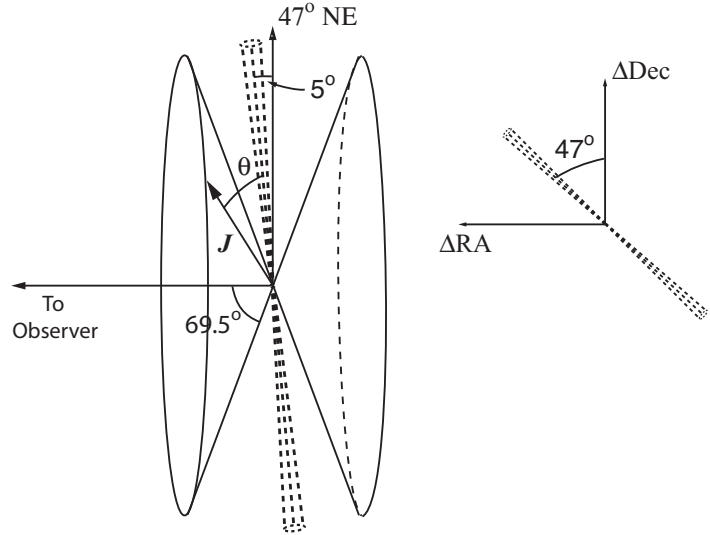}
\caption{Illustration of the relative orientation between the radio jets and the binary angular momentum axis, \boldmath$J$\unboldmath, of GRO J1655-40.  The observed orientation of the radio jets is shown by the {\it dashed} cones \citep{hje95}.  \boldmath$J~$\unboldmath is restricted to lie on either of two $69\fdg5$ cones about the line of sight of the observer \citep{oro97}, as shown by the {\it solid} cones.  The tilt angle, $\theta$, between the radio jets and \boldmath$J~$\unboldmath is between $15\fdg5$ and $90^\circ$.  \label{GROJ}}
\end{figure}

The observational constraints on the orientation of this system are summarized in Figure \ref{GROJ}.  \citet*{oro97} found the inclination of the binary orbit of this partially eclipsing system to be $i=69\fdg5 \pm0\fdg1$ (measured from the plane of the sky).  This means that the angular-momentum axis of the binary, \boldmath$J\mbox{\unboldmath$_{binary}$}$\unboldmath, is tilted by the same amount from the line-of-sight of an observer.  However, the orientation of the binary orbit (i.e. position angle) in the sky is not known.  Thus, \boldmath$J\mbox{\unboldmath$_{binary}$}~$\unboldmath is only constrained to lie somewhere on a $69\fdg5$ cone about the line-of-sight as shown in Figure \ref{GROJ}.  \citet*{hje95} observed two highly collimated relativistic jets expanding from opposite sides of this source during a period of hard X-ray activity \citep{har95}.  The position angle on the sky of the approaching jet was $47^\circ \pm 1^\circ$ and its inclination was $i=5^\circ \pm 2^\circ$ (i.e. almost in the plane of the sky).  This means that the axis through the center of the radio jets was tilted with respect to \boldmath$J\mbox{\unboldmath$_{binary}$}~$\unboldmath by between $15\fdg5$ and $90^\circ$ as illustrated in Figure \ref{GROJ}.  Depending on the model used to describe the radio jets, their orientation should reflect the angular momentum axis of either the BH or the inner accretion disk.  On the other hand, the orientation of the binary orbit should define the orientation of the outer accretion disk.  Altogether this implies that the outer accretion disk is probably tilted with respect to the equatorial plane of the black hole by between $15\fdg5$ and $90^\circ$.  This makes GRO J1655-40 a strong candidate for the occurrence of the Bardeen-Petterson effect.

In order to estimate the Bardeen-Petterson transition radius using equation \ref{RBP}, we need to know the mass and angular momentum of the BH.  Optical investigations of the binary give a mass of $M_{\mathrm{BH}}=6.7\pm 1.2 M_\odot$ \citep{oro97,sha99}.  Modeling of the X-ray spectra from this system \citep{sob99} gives a probable angular momentum of $a_* \approx 0.5$, with an upper limit of $a_* < 0.7$.  Other authors, however, give values as high as $a_* = 0.95$ \citep*{cui98}.  Moreover, the exact value of $a_*$ is not critical to our conclusions.  We simply require an upper limit, which we take as $a_* \lesssim 0.95$.  This gives us an upper limit on the transition radius of $R_\mathrm{BP} \lesssim 290 R_\mathrm{GR}$, based on our adopted constraints on the scaling parameter $A$.  The physical constraint that the transition region lie outside the innermost stable circular orbit gives us a firm lower limit on the transition radius of $R_\mathrm{BP} \gtrsim 6 R_\mathrm{GR}$ for a non- or slowly-rotating BH.  The resultant range is $6 R_\mathrm{GR} \lesssim R_\mathrm{BP} \lesssim 290 R_\mathrm{GR}$ or 60 km $\lesssim R_{\mathrm{BP}}\lesssim$ 2900 km.  We caution, however, that these estimates of $M_\mathrm{BH}$ and $a_*$ derive from models that have assumed a single continuous accretion disk,
whereas we have shown that the Bardeen-Petterson effect creates an
inner and outer disk component inclined relative to one another.
This point may require further analysis, however the size of the inner accretion disk is extremely small compared to the estimated size of the outer accretion disk [$4.2 \times 10^6$ km \citep*{oro97}], so that this correction is likely to be small.

The range of Keplerian frequencies corresponding to our derived range of $R_\mathrm{BP}$ is 1 Hz $\lesssim \nu_{Kep,\mathrm{BP}} \lesssim$ 300 Hz.  This range of frequencies is consistent with three of the four QPOs observed in GRO J1655-40 \citep{rem99b}: the variable 14-28 Hz QPO and the relatively stable 9 and 300 Hz QPOs.  However, identification of the 300 Hz QPO with the Bardeen-Petterson transition radius would require that the transition region lie just outside the innermost stable circular orbit, which would make it difficult for an inner, aligned disk to form.  It seems more likely that the Bardeen-Petterson effect might be associated with either the 9 Hz or the 14-28 Hz QPO.  This conclusion seems to be supported by the spectral modeling of \citet{rem99b}, who identified the 300 Hz and the 14-28 Hz QPOs with the power-law component of their model rather than the disk component.  Only the 9 Hz QPO appeared to have the X-ray spectral characteristics expected for a purely disk-based oscillation \citep{rem99b}.  Hence, we tentatively identify the 9 Hz QPO in GRO J1655-40 as perhaps the best potential observation of the Bardeen-Petterson transition radius in an X-ray binary.  However, there is an ongoing debate as to whether the 9 Hz QPO is defintely disk-related and the other QPOs are not.  Hence, it is premature to definitively exclude the 14-28 or the 300 Hz QPOs as possible candidates.

\section{Neutron Star X-ray Binaries} \label{LMXB sect}
For neutron-star X-ray binaries, the parameter space available is much narrower than for black-hole binaries.  Both the mass and the angular momentum of NSs are tightly constrained by theory and observation.  Models for the NS equation of state give the following upper limits (\citealt{sal94}, \citealt*{coo94c}): $M_{NS} \lesssim 2.6 M_\odot$ and $a_{*,NS} \lesssim 0.7$.  Observed radio pulsars in binary systems are all consistent with $1.35 M_\odot \le M_{NS} \le 1.45 M_\odot$ \citep{tho99}.  If we consider these upper limits of $M_{NS}$, $a_{*,NS}$, and the parameter $A$, then we can use equation \ref{fBP} to set a lower limit for the Bardeen-Petterson orbital frequency in NSs: $\nu_{Kep,\mathrm{BP}} \gtrsim$ 3 Hz.  The upper limit for the Bardeen-Petterson orbital frequency is fixed by the orbital frequency at the inner edge of the accretion disk.  For a putative 1.4 $M_\odot$ NS with $a_*=0.1$, the orbital frequency at the innermost stable circular orbit is $\approx$ 1700 Hz.  This range of frequencies ($3 - 1700$ Hz) is consistent with most of the QPOs observed in NS LMXBs.  In $Z$-sources, it is consistent with the normal/flaring and horizontal branch oscillations and kHz QPOs.  In atoll sources, it is consistent with the low frequency QPOs, the peaked noise components, and the kHz QPOs.

However, in order for the Bardeen-Petterson effect to be a consistent interpretation, we should expect all the QPOs associated with this effect to show similar properties.  Tentatively adopting the 9 Hz QPO in GRO J1655-40 as our most likely example, we can use it as a template with which to identify other Bardeen-Petterson generated QPOs.  Toward this end, this QPO has at least two noteworthy features:  It has a relatively low coherence parameter ($Q=\nu / \Delta \nu <3$) and its frequency appears to be independent of the count rate in the detector \citep{rem99b}, though this conclusion is based on only a few observations.  Another peculiar feature of the QPO is that it is only present during some observations but is mostly absent.  This transient behavior is presumably related to the accretion dynamics, e.g. in our model, whether or not clumps happen to be crossing the transition region during an observation.  Hopefully this aspect can be better understood with future hydrodynamic simulations.

Among NS LMXBs, low-coherency QPOs are fairly common.  However, the combination of low coherency and count-rate independence seems to be most consistent with the $\sim 100$ Hz peaked-noise or Lorentzian component.  This 100 Hz feature was first identified in 4U 1728-34 \citep{for98a}.  It has been extensively studied in SAX J1808.4-3658 by \citet{wij98b}.  These authors were also the first to identify a similar feature in 4U 0614+09 \citep[based on observations by][]{men97b} and 4U 1705-44 \citep[based on observations by][]{for98c}.  Similar features have also been found in 4U 1820-30 \citep*{wij99}, Terzan 2 \citep{sun00}, and GS 1826-24 \citep{sun00}.  These observations may imply that the 100 Hz feature is fairly common in LMXBs.  This might, however, become a concern if this feature proves to be too common to be consistent with the likely occurence of the Bardeen-Petterson effect.

Irrespective of all this, we argue that the Bardeen-Petterson mechanism should produce QPOs of similar frequency in all NS LMXBs since all NS systems are characteristically very similar.  We have already argued that the NS masses are similar ($M_{NS} \approx 1.4 M_\odot$).  Furthermore, interpretation of the nearly-coherent oscillations seen during Type I X-ray bursts in ten NS LMXBs argue for a narrow range of angular momenta for the NSs in these systems \citep[see, e.g.][]{str01}.  The two remaining parameters in the estimates of the Bardeen-Petterson transition radius, the viscosity and the disk aspect ratio, are presumably similar in all LMXBs since the accretion mechanism (Roche-lobe overflow) is assumed to be the same.  In this context, it is interesting that $\sim 100$ Hz QPOs may be quite common among NS LMXBs.  This frequency is also comfortably within the range (3-1700 Hz) expected for the Bardeen-Petterson effect in NSs, as calculated above.  Therefore, all of the parameters can assume typical values if the 100 Hz QPO is triggered by the Bardeen-Petterson effect.

For now, the association of the 100 Hz feature in NS LMXBs with the 9 Hz QPO in GRO J1655-40 is just a suggestion.  Other authors have suggested different associations.  For instance, based on a possible correlation between the 9 Hz and the 300 Hz QPO in GRO J1655-40, \citet*{psa99} suggest that these two QPOs are similar to the following QPO pairs seen in other systems:  the $\simeq 10-50$ Hz and the lower kHz QPOs in 4U 1728-34 and 4U 1608-52 and the $\sim 0.08-13$ Hz and $\sim 160-220$ Hz QPOs in XTE J1550-564.  However, these authors acknowledge that these identifications are very tentative and must be scrutinized carefully.

\section{Discussion}
Positive association of a QPO in a LMXB with the orbital frequency at the Bardeen-Petterson transition radius could provide important constraints on the mass and angular momentum of the accreting body and possibly the thickness and tilt of the accretion disk.  Careful study of the properties of such a QPO could also provide information about the gas dynamics near the transition radius and the role of the accretion rate in this model.  Furthermore, identification of such QPOs in several LMXBs could provide information about the relative abundance of tilted accretion disk systems, which is important in understanding their formation and evolution.

We have argued that the Bardeen-Petterson effect may be a common phenomenon in LMXBs and that it provides a plausible physical explanation for at least some observed quasi-periodic brightness oscillations.  The estimates of our model are consistent with a 1-300 Hz QPO in the black-hole X-ray binary GRO J1655-40.  This system is an important test since it appears likely to have a tilted accretion disk around a rapidly-rotating BH.  We have also argued that this mechanism could generate a moderate-frequency QPO when applied to individual neutron-star LMXBs.

We are currently undertaking the task of studying the Bardeen-Petterson effect numerically to follow the dynamics of the gas near the transition radius.  The results of this study should provide further insight into the validity of the model we have proposed here.

\begin{acknowledgements}
The authors wish to thank the referee for numerous helpful suggestions and Charles Bailyn for his help in clarifying the relative orientation of the radio jets and the binary orbit of GRO J1655-40.  P. C. F. would like to thank the Arthur J. Schmidt Foundation for
fellowship support at the University of Notre Dame.  This work was
supported in part by the National Science Foundation under grant
PHY-97-22086.
\end{acknowledgements}



\clearpage

\end{document}